\documentclass[a4paper,12pt]{article}
\pdfoutput=1

\usepackage{jheppub}
\usepackage[utf8]{inputenc}
\usepackage{amsmath,amssymb,amsthm}
\usepackage{physics,mathtools}
\usepackage{tikz,float}
\usepackage{bbm}
\usepackage{subcaption}
\usepackage{booktabs}

\usepackage{amssymb}
\usepackage{pifont}

\makeatletter
\def\@fpheader{\vspace{0.1mm}}
\makeatother
\newcommand{\bw}{\begin{widetext}}
\newcommand{\ew}{\end{widetext}}
\newcommand{\bea}{\begin{eqnarray}}
\newcommand{\eea}{\end{eqnarray}}
\newcommand{\be}{\begin{equation}}
\newcommand{\ee}{\end{equation}}
\newcommand{\bca}{\begin{cases}}
\newcommand{\eca}{\end{cases}}

\newcommand{\ben}{\begin{enumerate}}
\newcommand{\een}{\end{enumerate}}

\usetikzlibrary{patterns}
\makeatletter
\let\over\@@over
\makeatother

\title{{{Stringy resolution of bouncing singularities \\
}}}
\author{Ilija Buri\'c,}
\emailAdd{burici@tcd.ie}
\author{Ivan Gusev,}
\emailAdd{gusevi@tcd.ie}
\author{Andrei Parnachev}
\emailAdd{parnachev@maths.tcd.ie}

\affiliation{School of Mathematics and Hamilton Mathematics Institute, Trinity College, Dublin 2, Ireland}

\abstract{We study the near-horizon geometry of near-extremal NS5 branes, which admits an exact worldsheet description in terms of the $SL(2,\mathbb{R})_k/U(1)$ coset. In the supergravity limit, the analytically continued time-domain reflection amplitude exhibits a singularity associated with a null geodesic bouncing off the black-hole singularity, analogous to the bouncing singularities of boundary correlators in asymptotically AdS spacetimes. We show that finite-string-length effects render this amplitude an entire function of complex time, completely resolving the bouncing singularity. This resolution is nonperturbative in the inverse coset level: the singularity persists at every finite order in the expansion.}

\begin{document}

\maketitle

\section{Introduction}

Spacetime singularities in general relativity and their resolution through quantum effects continue to be some of the main motivations for studying quantum theories of gravity. In the context of the AdS/CFT correspondence, recent approaches have focused on the so-called bouncing geodesics, \cite{Fidkowski:2003nf,Festuccia:2005pi}. These are null geodesics that pass from one to the other boundary of a two-sided planar AdS Schwarzschild black hole, reaching all the way down to the black hole singularity in the bulk. The existence of bouncing geodesics leads to singularities in the complex time plane of the stress-tensor part of the thermal two-point function in the dual field theory, \cite{Ceplak:2024bja}, and, as a consequence, in the retarded correlator, \cite{Afkhami-Jeddi:2025wra}. Such singularities are often called bouncing singularities and they have been analyzed in several contexts and using various techniques, \cite{Parisini:2023nbd,Anegawa:2024kdj,Bajc:2025jjv,Ceplak:2025dds,Chakravarty:2025ncy,Jia:2025jbi,AliAhmad:2026wem,Jia:2026pmv,Araya:2026shz,Giombi:2026kdz,Arnaudo:2026der,Grozdanov:2026cut,Jia:2026ryl,Bajc:2026wzi}. Related thermal-bootstrap and double-trace analyses have also been developed, \cite{Buric:2025anb,Barrat:2025nvu,Buric:2025fye,Niarchos:2025cdg,Barrat:2025twb,Niarchos:2026wsw,Buric:2026cmn,Barrat:2026jfg,Guo:2026xyl,Correia:2026fnr}. Connections to high-overtone quasinormal spectra and near-singularity geometry have also been explored, \cite{Aniceto:2026zmc,Arnaudo:2026axe,Hartnoll:2026vhu,Xiao:2026pir,Ceplak:2026ard,Coviello:2026hdc}. Notably, these singularities persist in asymptotically flat and de Sitter spaces for suitable observables analogous to the boundary two-point function, \cite{Faruk:2025bed,Arnaudo:2026tcy,Grozdanov:2026ktq,Grozdanov:2026lnc}.
\smallskip

The fate of bouncing singularities beyond the classical limit was studied by analyzing simple quantum models, such as the SYK model, \cite{Dodelson:2024atp,Dodelson:2025jff,Dodelson:2026gak,Buric:2026qsp}. 
In these examples bouncing singularities were still present in the complex time plane.
Similarly, in ${\cal N}=4$ super Yang--Mills, keeping $N$ infinite and making the 't Hooft coupling $\lambda$ finite perturbatively seems to shift the singularities in the complex time plane, \cite{Ceplak_to_appear}.
\smallskip

In this work, we study the resolution of bouncing singularities in the asymptotically flat setup. In this case, in the absence of the asymptotic boundary, the correct observable to consider is the reflection amplitude $R(\omega)$\footnote{We comment on the relation to the retarded Green's function below.}. We begin by reviewing the computation of the large-$\omega$ reflection amplitude on the four-dimensional Schwarzschild black hole, \cite{Ben-Israel:2017zyi,Itzhaki:2018rld,Basha:2018bvi}. The large $\omega$ behavior is controlled by a two-legged null geodesic that comes from asymptotic infinity and bounces off the black hole singularity. The real time shift Re$(\Delta T)$ corresponding to this geodesic is ill-defined and depends on the cutoff of the asymptotic region. However, the imaginary part of the time shift is cutoff-independent and given by
\begin{equation}\label{intro-imT-beta}
    \text{Im}(\Delta T) = \frac{\beta}{2}\,,
\end{equation}
where $\beta$ is the inverse Hawking temperature. It is precisely the imaginary part that controls $R(\omega)$ at large $\omega$, giving
\begin{equation}\label{intro-Romega}
    R(\omega) \sim e^{-\frac{\beta \omega}{2}}\ .
\end{equation}
Upon taking the Fourier transform, this leads to a singularity of the reflection amplitude $R(t)$ at $t = i\beta/2$.
\smallskip

After reviewing the Schwarzschild case, we repeat the same analysis for the cigar geometry, which appears as part of the NS5 supergravity background in the near-horizon limit. The conclusions \eqref{intro-imT-beta} and \eqref{intro-Romega} remain unchanged, with the additional property that the reflection coefficient on the cigar can be written exactly in terms of gamma functions, see equation \eqref{R-cigar-classical-exact}.
\smallskip

The near-horizon limit of the NS5 supergravity appears as the classical limit of little string theory, when the number of branes $k$ goes to infinity. Consequently, the classical reflection amplitude \eqref{R-cigar-classical-exact} is the $k\to\infty$ limit of the $SL(2,\mathbb{R})_k/U(1)$ CFT reflection amplitude, which is known exactly, \cite{Teschner:1997ft,Teschner:1999ug,Giribet:2000fy,Giribet:2001ft}, see \eqref{reflection coefficient}. Despite their equality in the $k\to\infty$ limit, the two amplitudes have very different analytic structure. Indeed, we show that the CFT reflection amplitude admits an analytic continuation to an entire function in the time domain. Therefore, the bouncing singularity gets resolved by worldsheet effects.
\smallskip

The paper is organized as follows. In Section \ref{S:Reflection amplitudes and bouncing geodesics}, we compute the reflection amplitude on the Schwarzschild and cigar backgrounds and relate their large-$\omega$ behavior to bouncing geodesics. In Section \ref{S:Reflection amplitude in little string theory}, we review the amplitude in the $SL(2,\mathbb{R})_k/U(1)$ CFT and show that it gives an entire function in the time domain. We  discuss our results in Section \ref{S:Discussion}.
Appendix A discusses the relation between the reflection amplitude and the retarded Green's function. 
Appendix B contains details of computing the reflection amplitude.

\medskip
While this paper was being finalized, \cite{Komargodski:2026oob} and \cite{Batra:2026gbq} appeared, addressing the resolution of bouncing singularities from different points of view.

\section{Reflection amplitudes and bouncing geodesics}
\label{S:Reflection amplitudes and bouncing geodesics}

In this section, we discuss the reflection amplitude on two backgrounds: the four-dimensional Schwarzschild black hole and the two-dimensional cigar geometry. The reflection amplitude is defined in terms of solutions to a certain wave equation on the background. In the case of Schwarzschild, this will be the Klein-Gordon (KG) equation. For the cigar, the KG operator will be dressed by the dilaton field, whose origin comes from the supergravity solution that the cigar background is a part of. In both cases, we will first find ‘bouncing' radial null geodesics. The imaginary part of the time shift associated with these geodesics controls the large-$\omega$ asymptotics of the reflection amplitude $R(\omega)$.

\subsection{Schwarzschild black hole}

Consider the four-dimensional Schwarzschild black hole
\begin{equation}\label{Schwarzschild-metric}
   ds^2 = -\left(1-\frac{r_h}{r}\right) dt^2 + \frac{dr^2}{1-\frac{r_h}{r}} + r^2 d\Omega_2^2\,,
\end{equation}
where $r_h = 2M_{\rm Schw}$ denotes the Schwarzschild radius, $M_{\rm Schw}$ is the mass and we have set the Newton constant to $G_N=1$. The $d\Omega_2^2$ denotes the round metric on $S^2$. The inverse temperature is given by
\begin{equation}
\beta = 4\pi r_h = 8\pi M_{\rm Schw}\ .
\end{equation}
We shall denote the tortoise coordinate by
\begin{equation}
r_*(r) = r + r_h\log\left(\frac{r}{r_h}-1\right)\ .
\end{equation}
We complexify the coordinate $r$ and choose a branch for the logarithm so that at the singularity $r=0$ we have
\begin{equation}
r_*(0) = r_h\log(-1) \equiv i\pi r_h\ .
\end{equation}

\paragraph{Null geodesic.} Consider a two-legged null geodesic from a radial cutoff $r=r_c$ to the singularity and back. We denote
\begin{equation}
\Delta T_{\rm Schw}(r_c) \equiv 2\left[r_*(0) - r_*(r_c)\right] = 2\pi i\, r_h - 2r_c - 2r_h\log\left(\frac{r_c}{r_h}-1\right)\ .
\end{equation}
The real part is cutoff-dependent and diverges as $r_c\to\infty$. The imaginary part is cutoff-independent and reads
\begin{equation}\label{DeltaT}
\operatorname{Im}\Delta T_{\rm Schw} = 2\pi r_h = \frac{\beta}{2}\ .
\end{equation}

\paragraph{Reflection amplitude.} Let us turn to the reflection amplitude. To this end, we consider a solution to the massless wave equation on the background \eqref{Schwarzschild-metric} in the separated form
\begin{equation}
    \Phi = \frac{1}{r}\, \psi_{\omega\ell}(r)\, e^{-i\omega t}\, Y_{\ell m}(\theta,\varphi)\,,
\end{equation}
where $(\theta,\varphi)$ are coordinates on $S^2$ and $Y_{\ell m}$ the usual spherical harmonics. The equation for $\psi_{\omega\ell}$ takes the Schr\"odinger form
\begin{equation}\label{KG-equation-Schwarzshild}
    \left( \partial_{r_*}^2 + \omega^2 - V_\ell\right) \psi_{\omega\ell} = 0\,,
\end{equation}
with the potential $V_\ell$ given by
\begin{equation}
    V_\ell = \left( 1 - \frac{r_h}{r} \right) \left( \frac{\ell(\ell+1)}{r^2} + \frac{r_h}{r^3} \right)\ .
\end{equation}
We regard this as a scattering problem in the region outside of the horizon, $r_h < r < \infty$. In this domain, the tortoise coordinate $r_*$ ranges over the real line, $r_*\in(-\infty,\infty)$, and the potential $V_\ell$ vanishes at the two boundaries. Therefore, the near-boundary solutions reduce to plane waves. We look for the solution that is purely ingoing at the horizon
\begin{equation}\label{boundary-conditions-Schwarzschild}
    \psi_{\omega\ell} \sim \begin{cases}
    T_\ell(\omega) e^{-i\omega r_*}\,, \hskip2.1cm r_* \to -\infty \\
   e^{-i\omega r_*} + R_\ell(\omega) e^{i\omega r_*}\,, \qquad r_* \to \infty
\end{cases}\ .
\end{equation}
We are interested in the reflection amplitude $R_\ell(\omega)$ at large $\omega$ and finite $\ell$. Using the first Born approximation, we have (see e.g. \cite{Ben-Israel:2017zyi,Basha:2018bvi,Itzhaki:2018rld})\footnote{We use $a\sim b$ to denote agreement of leading exponential behaviors of $a$ and $b$ as $\omega\to\infty$ and $a\simeq b$ to mean $a/b\to 1$ as $\omega\to\infty$.}
\begin{equation}
     R_\ell(\omega) \sim \frac{1}{2i\omega} \int_{-\infty}^\infty dr_*\, e^{-2i\omega r_*} V_\ell \sim e^{-2\pi r_h \omega}\,, \qquad \omega\to\infty\ .
\end{equation}
At large $\omega$, the integral is dominated by the singularity of $V_\ell(r_*)$ closest to the real $r_*$--axis. At this point we have
\begin{equation}
     r_\ast = r_*^{\rm sing} \equiv - i \pi r_h = -\frac{i\beta}{4}\ .
\end{equation}
Computing the residue, we find
\begin{equation}\label{Schwarzschild-exponential-falloff}
     R_\ell(\omega) \sim e^{-2i\omega r_*^{\rm sing}} = e^{-\frac{\beta\omega}{2}}\ .
\end{equation}
The above computation does not determine the prefactor multiplying the exponential, as the latter requires a more detailed analysis. We will only be interested in the coefficient of the exponential decay and for that purpose \eqref{Schwarzschild-exponential-falloff} is sufficient. We see that the decay coefficient is equal to Im$(\Delta T_{\rm Schw})$ obtained in \eqref{DeltaT}. This is a more general phenomenon and we shall see it in another example below. Equations \eqref{DeltaT} and \eqref{Schwarzschild-exponential-falloff} are the main takeaway points from this subsection.
\smallskip

Finally, let us point out that our main conclusions persist in the case of massive fields. Namely, for a field of mass $M$, the equation \eqref{KG-equation-Schwarzshild} gets modified by changing the potential $V_\ell\to V_{\ell,M}$, where
\begin{equation}\label{Schwarzschild-potential-massiave}
     V_{\ell,M} = \left( 1 - \frac{r_h}{r} \right) \left(M^2 + \frac{\ell(\ell+1)}{r^2} + \frac{r_h}{r^3} \right)\ .
\end{equation}
The potential vanishes $V_{\ell,M}\to0$ at the horizon and tends to a constant $V_{\ell,M}\to M^2$ at the asymptotic boundary. Therefore, the boundary conditions for the corresponding scattering problem are

\begin{equation}\label{boundary-conditions-Schwarzschild-massive}
    \psi_{\omega\ell,M} \sim \begin{cases}
    T_{\ell,M}(\omega) e^{-i\omega r_*}\,, \hskip2.1cm r_* \to -\infty \\
   e^{-i p r_*} + R_{\ell,M}(\omega) e^{i p r_*}\,, \qquad r_* \to \infty
\end{cases}\,,
\end{equation}
where $p^2 = \omega^2 - M^2$. At large $\omega$, we have $p\simeq\omega$ and the Born approximation analysis proceeds as before,
\begin{equation}
     R_{\ell,M}(\omega) \sim \frac{1}{2i\omega} \int_{-\infty}^\infty dr_*\, e^{-2i\omega r_*} V_{\ell,M}(r_*) \sim e^{-\frac{\beta\omega}{2}}\,, \qquad \omega\to\infty\ .
\end{equation}
The integral is dominated by the pole of $V_{\ell,M}$ at $r_* = r_*^{\rm sing}$, giving the same leading exponential behavior of the reflection amplitude as in \eqref{Schwarzschild-exponential-falloff}.

\subsection{Near-extremal NS5 branes}

In this subsection, we repeat the above analysis on the cigar geometry, which appears as part of the NS5 supergravity background in the near-horizon limit. After reviewing the full NS5 geometry and the relevant supergravity fields, we will focus on the two-dimensional cigar factor. We will compute the reflection amplitude for a minimally coupled scalar field and relate it to the bouncing null geodesic.

\paragraph{Supergravity background.} Little string theory on NS5-branes is defined by their decoupling limit \cite{Seiberg:1997zk}, and its holographic description involves string theory in the near-horizon background \cite{Aharony:1998ub}. We start from the non-extremal NS5 supergravity background, whose metric (written in the string frame), dilaton and $B$-fields are
\begin{align}
    & ds^2_{\rm string} = - f(r) d \tilde t^2 + A(r) \left(\frac{dr^2}{f(r)} + r^2 d\Omega_3^2\right) + dx_5^2\,,\\
    & e^{2\Phi_D} = g_s^2 A(r)\,, \qquad H_3 = 2 \left(k\alpha'\right)^\frac12 \sqrt{r_0^2 + k\alpha'} \epsilon_3\,,
\end{align}
where
\begin{equation}
    f(r) = 1 - \frac{r_0^2}{r^2}\,, \qquad A(r) = 1 + \frac{k\alpha'}{r^2}\ .
\end{equation}
Here, $dx_5^2$ is the flat metric on the five-dimensional Euclidean space corresponding to directions along the branes and $\Omega_3,\epsilon_3$ denote the metric and the volume form of the round unit three-sphere in the space transverse to the brane. The metric of the transverse space has a horizon at $r=r_0$ and $k$ denotes the number of branes. We are interested in the near-horizon decoupling limit \cite{Horowitz:1991cd,Maldacena:1997cg}, which is defined by taking $r_0,g_s\to0$, while keeping the quantity
\begin{equation}
    \mu = \frac{r_0^2}{g_s^2 \alpha'}\,,
\end{equation}
finite. In this limit, the metric and the dilaton become
\begin{align}\label{near-horizon-metric}
    & ds^2_{\rm string} = k\alpha' \left( d\phi^2 - \tanh^2\phi\, dt^2 + d\Omega_3^2 \right) + dx_5^2\,,\\
    & \Phi_D = \frac12\log\frac{k}{\mu} - \log\cosh\phi \equiv \Phi_0 - \log\cosh\phi \,,\label{near-horizon-dilaton}
\end{align}
where we have introduced 
\begin{equation}
    r = r_0 \cosh\phi\,, \qquad \tilde t = \sqrt{k\alpha'}\, t \ .
\end{equation}
The dimensionless Euclidean time $\tau = it$ has period $2\pi$. In the near-horizon limit, the geometry is $\mathbb{R}^5 \times S^3 \times \mathcal{M}_{\rm cigar}$, where the cigar metric is given by
\begin{equation}\label{cigar-metric}
    ds_2^2 = k\alpha' \left( d\phi^2 - \tanh^2\phi\, dt^2 \right)\ .
\end{equation}
The inverse temperature associated to the cigar geometry, determined by requiring absence of a conical singularity in Euclidean signature, reads
\begin{equation}
 \beta = 2\pi\sqrt{k\alpha'}\ .
\end{equation}

\paragraph{Null geodesic.} We focus on the cigar part \eqref{cigar-metric} of the NS5 geometry. The geometry has a horizon at $\phi=0$, the exterior at $\phi>0$ and the singularity is reached by analytic continuation
\begin{equation}
    \phi=i\theta\,,
\end{equation}
and sits at $\theta = \pi/2$. Inside the horizon, we have
\begin{equation}
 ds_2^2 = k\alpha' \left( -d\theta^2 + \tan^2\theta\, dt^2 \right)\ .
\end{equation}
The null tortoise coordinate is defined as
\begin{equation}\label{null-tortoise}
z = \int^\phi \frac{d\phi'}{\tanh\phi'} = \log\sinh\phi\ .
\end{equation}
Therefore, the asymptotic boundary is at $z\to\infty$, the horizon at $z\to-\infty$ and the singularity at
\begin{equation}\label{singularity}
z_s = \log\sinh\left(\frac{i\pi}{2}\right) = \log i = \frac{i\pi}{2}\ .
\end{equation}
In the last line, we have chosen a branch for the logarithm. Null geodesics on the cigar satisfy
\begin{equation}
   d\phi^2-\tanh^2\phi\,dt^2 = 0\ .
\end{equation}
Integrating gives
\begin{equation}
   t = \pm \log\sinh\phi+\text{const} = \pm z+\text{const}\ .
\end{equation}
Let $\phi_c$ be a radial cutoff and $z_c$ the corresponding tortoise value, $z_c=\log\sinh\phi_c$. A one-sided null geodesic from the cutoff to the singularity \eqref{singularity} has the complex time difference
\begin{equation}
\Delta t_{\rm leg} = z_s - z_c = \frac{i\pi}{2}-z_c\ .
\end{equation}
In direct analogy with the Schwarzschild case \eqref{DeltaT}, we denote twice this value by
\begin{equation}
\Delta t_{\rm null}(\phi_c) = 2 \Delta t_{\rm leg} = i\pi-2z_c = i\pi-2\log\sinh\phi_c\ .
\end{equation}
As for the Schwarzschild black hole, the real part is cutoff-dependent and diverges as $\phi_c\to\infty$, while the imaginary part is cutoff-independent and reads
\begin{equation}
\operatorname{Im}\Delta t_{\rm null} = \pi\ .
\end{equation}
In physical units, we get
\begin{equation}
\operatorname{Im}\Delta \tilde t = \pi\sqrt{k\alpha'} = \frac{\beta}{2}\ .
\end{equation}

\paragraph{Reflection amplitude.} We turn to the reflection amplitude. Let $G^{\rm string}$ and $G^E$ denote the ten-dimensional NS5 metric in the string frame and in the Einstein frame, respectively. We have
\begin{equation}
    G^{\rm string}_{AB} = e^{\frac12 \Phi_D} G^E_{AB}\,, \qquad A,B = 0,\dots, 9\ .
\end{equation}
We consider a massless, minimally coupled scalar field $\Psi$ with the action
\begin{equation}
    S = \int d^{10}x \sqrt{|G_E|}\, G_E^{AB} \partial_A\Psi \partial_B\Psi = \int d^{10}x\, e^{-2\Phi_D}\, \sqrt{|G_{\rm string}|}\, G_{\rm string}^{AB} \partial_A\Psi \partial_B\Psi\,,
\end{equation}
and focus on the case where the field depends non-trivially only on the two coordinates of the cigar geometry, $\Psi = \Psi(t,\phi)$. Notice that $\Psi$ is minimally coupled with respect to the Einstein frame metric, rather than the string frame metric. This means that the equation of motion involves the KG operator dressed by factors of the dilaton,
\begin{equation}\label{dilaton-dressed-KG-equation}
    \frac{e^{2\Phi_D}}{\sqrt{-g}}\partial_\mu \left(\frac{\sqrt{-g}}{e^{2\Phi_D}} g^{\mu\nu}\partial_\nu \Psi\right) = 0\,,
\end{equation}
where $g$ is the metric \eqref{cigar-metric}. The analysis of the reflection amplitude proceeds similarly as in the Schwarzschild case. We look for separated solutions of the form
\begin{equation}
    \Psi = e^{-i\hat\omega t} \frac{\psi(z)}{\cosh\phi}\,,
\end{equation}
for which the equation \eqref{dilaton-dressed-KG-equation} reduces to
\begin{equation}\label{Schroedinger-eqn-cigar}
    \left( \partial_z^2 + \hat\omega^2 - V\right) \psi = 0\,, \qquad V = 1 - \frac{1}{\left(1+e^{2z}\right)^2}\ .
\end{equation}
The equation is solved in the region $z\in(-\infty,\infty)$. On the two boundaries, the potential behaves as $V\to0$ (at $z\to-\infty$) and $V\to1$ (at $z\to\infty$). Therefore, we look for a solution with the asymptotic behavior
\begin{equation}\label{cigar-boundary-conditions}
    \psi_{\hat\omega} \sim \begin{cases}
T(\hat\omega) e^{-i\hat\omega z}\,, \hskip1.9cm z \to -\infty \\
e^{-i \hat q z} + R(\hat\omega) e^{i \hat q z}\,, \qquad z \to \infty
\end{cases}\,,
\end{equation}
where $\hat q^2 = \hat\omega^2 - 1$. The solution at large $\hat\omega$ may be obtained similarly as for the Schwarzschild black hole. The integral in the Born approximation is dominated by the pole of $V(z)$ at $z = - z_s$. However, for the cigar geometry, the exact reflection amplitude can in fact be written in terms of the gamma functions, \cite{Itzhaki:2018rld}. Indeed, the exact solution to \eqref{cigar-boundary-conditions} is given by
\begin{equation}
    \psi_{\hat\omega} = T(\hat\omega)\, y^{\frac{i\hat q}{2}} (1 - y)^{-\frac{i\hat\omega}{2}}\ _2F_1\left(\frac{1+i\hat q - i\hat\omega}{2},\frac{1+i\hat q - i\hat\omega}{2},1 - i\hat\omega,1-y\right)\,,
\end{equation}
where $y = 1/(1 + e^{2z}) = \cosh^{-2}\phi$. In the last line, $T(\hat\omega)$ is an arbitrary normalization constant that we have denoted so that it agrees with \eqref{cigar-boundary-conditions}. One reads off the reflection amplitude
\begin{equation}\label{R-cigar-classical-exact}
    R(\hat\omega) = \frac{\Gamma(i \hat q)}{\Gamma(-i \hat q)} \frac{\Gamma\left(\frac{1-i\hat q-i\hat\omega}{2}\right)^2}{\Gamma\left(\frac{1+i\hat q-i\hat\omega}{2}\right)^2} \equiv R_{\rm classical}^{\rm exact}(\hat\omega)\ .
\end{equation}
We attached the index ‘classical' to the above amplitude to contrast it to a CFT reflection amplitude analyzed in the next section. Expanding at large $\hat\omega$, we obtain
\begin{equation}\label{R-cigar-classical-leading}
    R(\hat\omega) \simeq -2i e^{-\pi\hat\omega} = -2i e^{-\pi \sqrt{k\alpha'}\omega} = -2i e^{-\frac{\beta\omega}{2}}\,,
\end{equation}
where we have introduced the dimensionful frequency $\omega = \hat\omega/\sqrt{k\alpha'}$. The exponential falloff of \eqref{R-cigar-classical-leading} is a direct analogue of the Schwarzschild result \eqref{Schwarzschild-exponential-falloff}. Finally, let us comment that the asymptotic behavior \eqref{R-cigar-classical-leading} leads to the pole of the Fourier transform of $R(\hat\omega)$ at the complex time $t = i\pi$,\footnote{The exact reflection coefficient $R(t)$ and the lattice of its singularities at $t_m=i\pi(2m+1)$ can be found in Appendix \ref{app:sugra-analytic-structure}.}
\begin{equation}
    R(t) = \int^\infty \frac{d\hat\omega}{2\pi}\, e^{-i\hat\omega t} R(\hat\omega) \sim -\frac{1}{\pi} \frac{1}{t - i\pi}\ .
\end{equation}
The corresponding dimensionful time is
\begin{equation}\label{bouncing-singularity-cigar}
   \tilde t_{\rm sing} = i\pi \sqrt{k\alpha'} = \frac{i\beta}{2}\ .
\end{equation}
In analogy with asymptotically AdS cases, \cite{Fidkowski:2003nf,Ceplak:2024bja}, we refer to this singularity as the bouncing singularity. Equations \eqref{R-cigar-classical-exact}, \eqref{R-cigar-classical-leading} and \eqref{bouncing-singularity-cigar} are the main results of this section.

Finally, let us note that the case of massive fields is treated similarly and leads to the same conclusions. Repeating the analysis, one gets
\begin{equation}\label{R-cigar-classical-exact-massive}
    R_{\rm classical}^{\rm exact}(\hat\omega, \hat M)= \frac{\Gamma(i \hat q_M)}{\Gamma(-i \hat q_M)} \frac{\Gamma\left(\frac{1-i\hat q_M-i\hat\omega}{2}\right)^2}{\Gamma\left(\frac{1+i\hat q_M-i\hat\omega}{2}\right)^2} \,,
\end{equation}
where the physical mass is $M = \hat M / \sqrt{k\alpha'}$ and the only difference to \eqref{R-cigar-classical-exact} lies in the dispersion relation, $\hat q^2_M = \hat\omega^2 - 1 -\hat M^2$. Since at large $\hat\omega$ we again have $\hat q_M \simeq \hat\omega$, the formulas \eqref{R-cigar-classical-leading} and \eqref{bouncing-singularity-cigar} are not changed.

\section{Reflection amplitude in little string theory}
\label{S:Reflection amplitude in little string theory}

The near-horizon limit of the NS5 geometry studied in the last section arises in the classical limit of little string theory. In this section, we analyze the reflection amplitude in the full worldsheet theory. It will turn out that the bouncing singularity of the classical reflection amplitude gets resolved by worldsheet effects.

\subsection{Little string theory} 

We begin by briefly reviewing those facts about little string theory that we will need in the analysis below. The cigar factor is described by the supersymmetric version of the two-dimensional black-hole coset \cite{Witten:1991yr}. The full near-horizon background is described by the worldsheet superconformal field theory with the target space $\mathbb{R}^5 \times SL(2,\mathbb{R})_k/U(1) \times SU(2)_k$. We denote the CFT operators schematically as
\begin{equation}
    \mathcal{O}_{\rm CFT} \sim e^{ip \cdot x}\, V_{j;m,\bar m}\, \Phi_{\ell;m',\bar m'} \times \text{fermions} \ .
\end{equation}
Here, $x$ denotes components of the free boson and $V_{j;m,\bar m}$ and $\Phi_{\ell;m',\bar m'}$ are primaries of the bosonic $SL(2,\mathbb{R})/U(1)$ coset and $SU(2)$ WZNW models, of levels $k\pm2$, respectively. The supersymmetric CFT contains additional free fermions and operators in little string theory also have contributions from ghosts that will not play a role in what follows. The three factor SCFTs have central charges
\begin{equation}
    c_{\mathbb{R}^5} = \frac{15}{2}\,, \qquad c_{\rm cigar} = 3 + \frac{6}{k}\,, \qquad c_{SU(2)} = \frac92 - \frac{6}{k}\ .
\end{equation}
In most of our analysis, we will focus on the field $V_{j;m,\bar m}$, which is a primary of the $SL(2,\mathbb{R})_k/U(1)$ coset model, with holomorphic and anti-holomorphic scaling dimensions
\begin{equation}\label{scaling-dimensions-coset}
    \Delta_{j;m,\bar m} = \frac{m^2 - j(j+1)}{k}\,, \qquad \bar\Delta_{j;m,\bar m} = \frac{\bar m^2 - j(j+1)}{k}\ .
\end{equation}
The other factors of the CFT will enter the analysis only through the physical state condition
\begin{equation}\label{physical-state-conditions}
   \frac{\alpha'}{4} p^2 + \Delta_{j;m,\bar m} +\frac{\ell(\ell+1)}{k} + N = \frac{\alpha'}{4} p^2 + \bar\Delta_{j;m,\bar m} +\frac{\ell(\ell+1)}{k} + \bar N = \frac12\ .
\end{equation}
Since in the supergravity limit we consider (massless) excitations only along the cigar (and in the NS-NS sector), we may consistently restrict to the sector $p=0$, $\ell=0$, $N = \bar N = 1/2$. Furthermore, for the rest of this section, we assume that $m = -\bar m\equiv n/2$, i.e. we consider states with zero winding around the asymptotic circle in the cigar geometry. We will denote the corresponding operators by $V_{j,n}$. The conditions \eqref{physical-state-conditions} become
\begin{equation}\label{simplified-physical-state-condition}
   \frac{n^2}{4} -j(j+1) = 0\ .
\end{equation}

\subsection{Reflection coefficient}

In this section, we discuss the exact reflection coefficient in the $SL(2,\mathbb R)_k/U(1)$ coset CFT describing the cigar geometry, see also \cite{Aharony:2004xn}. Related thermal two-point functions in the near-horizon NS5 background and their supergravity limits were studied in \cite{Narayan:2001dr,DeBoer:2003dd,Parnachev:2005hh}. In the asymptotic linear-dilaton region, $\phi\to\infty$, a Virasoro primary with angular momentum $n\in\mathbb Z$ behaves as
\begin{equation}
V_{\alpha,n}\sim e^{in\tau}e^{\alpha\phi}\,,
\end{equation}
where $\tau\sim\tau+2\pi$ parametrizes the asymptotic circle and $\alpha$ determines the radial dependence. Its chiral conformal dimension is given by
\begin{equation}
     \Delta = \frac{n^2}{4k}-\frac{1}{2}\alpha(\alpha+Q)\,,
\end{equation}
where $Q^2=2/k$. Comparing with the scaling dimension \eqref{scaling-dimensions-coset}, we find two solutions for $\alpha$ in terms of $j$,
\begin{equation}
\alpha=Qj \quad\text{or}\quad \alpha=-Q(j+1)\ .
\end{equation}
Thus, the two radial asymptotic behaviors correspond to $j$ and $-j-1$, which have the same conformal dimension. The asymptotic form of the coset primary can therefore be written as
\begin{equation}
V_{j,n}\sim e^{in\tau} \left[e^{Qj\phi} + R(j,n)e^{-Q(j+1)\phi}\right]\,, \qquad \phi\to\infty\,,
\end{equation}
where we have normalized the coefficient of the first term to unity. The relative coefficient $R(j,n)$ is the reflection coefficient/amplitude. The latter can be computed exactly in the CFT using either conformal bootstrap or screening-charge methods \cite{Teschner:1999ug,Giribet:2000fy,Giribet:2001ft}. Specializing to the zero-winding sector gives \cite{Giveon:1999px,Giveon:1999tq}\footnote{Here $n\in\mathbb Z$ is the Euclidean momentum around the cigar. Therefore, \eqref{reflection coefficient} determines the reflection coefficient only at a discrete set of frequencies and does not, by itself, specify a unique analytic continuation to complex $n$. For example, one may add $\sin(\pi n)\,F(j,n)$ for any suitable function $F(j,n)$ without changing the Euclidean answer. Uniqueness requires additional analyticity and growth conditions, together with a choice of Lorentzian boundary conditions \cite{BaymMermin:1961,Cuniberti:2001hm}. Following usual practice, we will regard \eqref{reflection coefficient} as the amplitude at any complex value of $n$.}

\begin{equation}\label{reflection coefficient}
R(j,n) = \nu(k)^{2j+1}
\frac{\Gamma\left(1-\frac{2j+1}{k}\right)}{\Gamma\left(1+\frac{2j+1}{k}\right)}
\frac{\Gamma(-2j-1)}{\Gamma(2j+1)}\frac{\Gamma\left(j+1+\frac{n}{2}\right)^2}{\Gamma\left(-j+\frac{n}{2}\right)^2}\ .
\end{equation}
This coefficient encodes the reflection of the asymptotic radial wave from the interior of the cigar. The normalization constant $\nu(k)$ is ambiguous and corresponds to shifting $\phi\to\phi+\phi_0$ in \eqref{reflection coefficient}. To compare with the classical solution \eqref{R-cigar-classical-exact}, we set
\begin{equation}\label{labels-nj-wq}
    n=-i\hat\omega\,, \qquad 2j+1=-i\hat q\ .
\end{equation}
Then the physical state condition \eqref{simplified-physical-state-condition} reproduces the relation $\hat q^2 = \hat\omega^2-1$. Substituting \eqref{labels-nj-wq} into \eqref{reflection coefficient}, we obtain
\begin{equation}
    R(j,n) \equiv R^{\rm exact}(\hat\omega) = \nu(k)^{-i\hat q}
\frac{\Gamma\left(1+\frac{i\hat q}{k}\right)}
{\Gamma\left(1-\frac{i\hat q}{k}\right)}
\frac{\Gamma(i\hat q)}
{\Gamma(-i\hat q)}
\frac{\Gamma\left(\frac{1-i\hat q-i\hat\omega}{2}\right)^2}
{\Gamma\left(\frac{1+i\hat q-i\hat\omega}{2}\right)^2}\ .
\end{equation}
The last two factors give precisely the classical result \eqref{R-cigar-classical-exact}, i.e.
\begin{equation}\label{quantum-classical-reflection-relation}
R^{\rm exact}(\hat\omega)
    =
 \nu(k)^{-i\hat q}
    \frac{\Gamma\left(1+\frac{i\hat q}{k}\right)}
    {\Gamma\left(1-\frac{i\hat q}{k}\right)}
    R_{\rm classical}^{\rm exact}(\hat\omega)\ .
\end{equation}
As $k\to\infty$, the ratio of gamma functions approaches one. The remaining factor $\nu(k)^{-i\hat q}$ is a convention-dependent phase associated with the choice of origin of the asymptotic radial coordinate. After matching this convention to the classical calculation, we recover
\begin{equation}
    R^{\rm exact}(\hat\omega) \longrightarrow R_{\rm classical}^{\rm exact}(\hat\omega)\,, \qquad k\to\infty\ .
\end{equation}

\paragraph{Remark.} The analysis for massive excitations is similar. In this case, one takes $N,\bar N>1/2$, the relation to the classical mass being
\begin{equation}
    \hat M^2 = 4k\left( N - \frac12\right)\ .
\end{equation}
Upon substitutions
\begin{equation}\label{labels-nj-wq-massive}
    n=-i\hat\omega\,, \qquad 2j+1=-i\hat q_M\,,
\end{equation}
the physical state conditions \eqref{physical-state-conditions} recover the relation $\hat q_M^2 = \hat\omega^2 - 1 - \hat M^2$.

\subsection{Bouncing singularity resolution}

Finally, let us show that the Fourier transform of the exact reflection coefficient \eqref{reflection coefficient} admits analytic continuation to an entire function in complex time. We consider
\begin{equation}\label{R(t)-Fourier}
    R(t) =
    \frac{1}{2\pi} \int_{-\infty}^\infty d\hat\omega\, e^{-i\hat\omega t} R^{\rm exact}(\hat\omega+i\epsilon)\ .
\end{equation}
The physical sheet is specified by
$\hat q=\sqrt{\hat\omega^2-1}\sim\hat\omega$ at infinity, and the boundary value is taken above the square-root cut.\footnote{The $+i\epsilon$ prescription is the usual retarded radiation prescription for the scattering modes.} Throughout this subsection, we work at fixed $k>1$ and choose
$\nu(k)=1$. On the physical sheet, the boundary value $R^{\rm exact}(\hat\omega+i\epsilon)$ is locally bounded on the real frequency axis. Indeed, it approaches $-1$ at $\hat\omega=\pm1$, while the apparent singularity at $\hat\omega=0$ is removable, with
\begin{equation}
    R^{\rm exact}(0+i\epsilon)
    =
    \frac{1}{2}\,
    \frac{\Gamma(1-1/k)}{\Gamma(1+1/k)}\ .
\end{equation}
The remaining gamma function factors introduce no other real frequency poles for $k>1$. Consequently, the contribution to \eqref{R(t)-Fourier} from any bounded frequency interval is an entire function of $t$. Establishing entireness of the full Fourier transform therefore reduces to controlling its two large-frequency tails $\hat\omega\to\pm\infty$.

\begin{figure}[t]
    \centering
    \includegraphics[width=\linewidth]{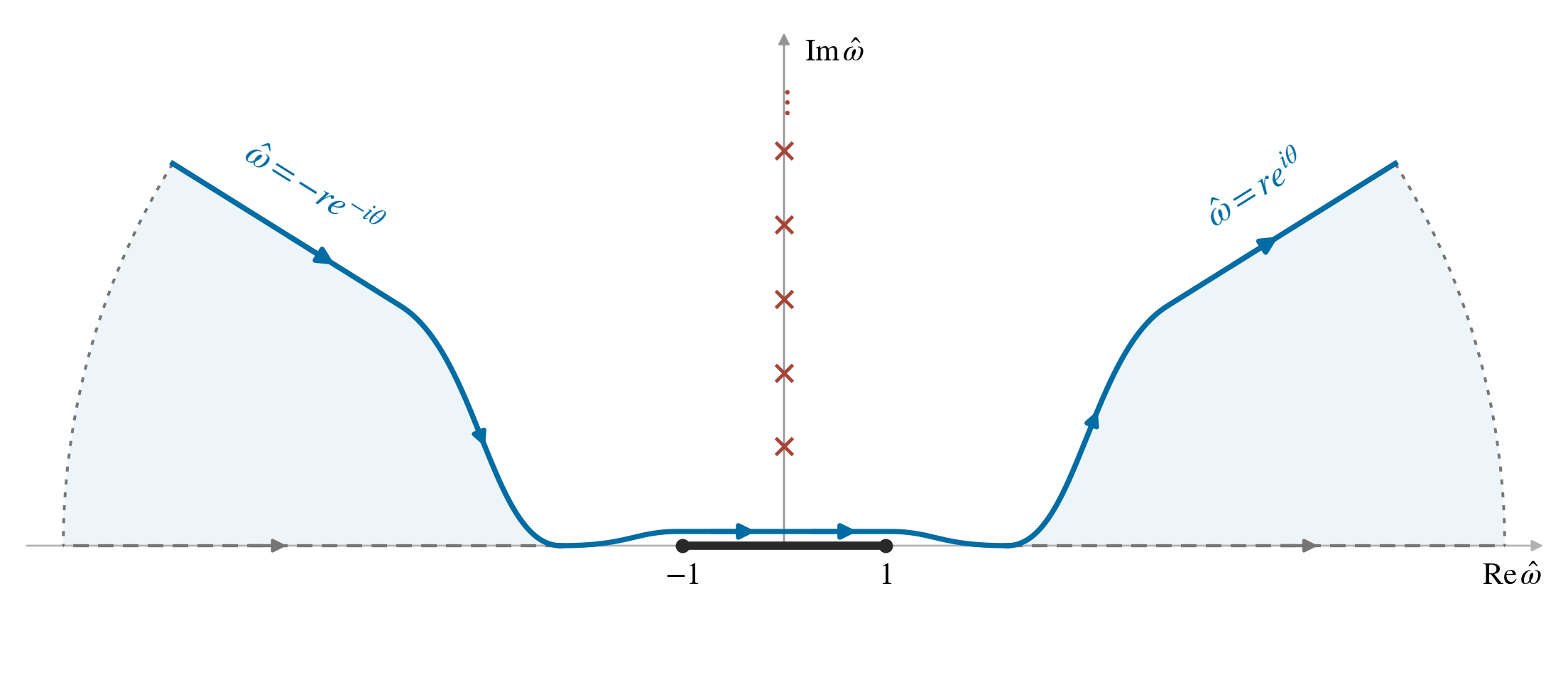}
    \caption{Contour deformation for the two-sided Fourier transform of $R^{\rm exact}(\hat\omega+i\epsilon)$. The blue contour approaches $\hat\omega=-r e^{-i\theta}$ on the left and $\hat\omega=r e^{i\theta}$ on the right as $r\to\infty$, with $0<\theta<\pi/2$. The finite-frequency part is unchanged; its displacement above the cut $[-1,1]$ represents the $+i\epsilon$ prescription. The dotted closing arcs vanish as their radius tends to infinity in the initial strip $-\pi<\operatorname{Im}t<\pi$. Along both rotated tails, the exact Fourier kernel decays faster than exponentially.}
    \label{fig:exact-reflection-contour}
\end{figure}

On the original real contour, the exponential decrease of the reflection coefficient \eqref{reflection coefficient} gives absolute convergence in the strip
\begin{equation}
    -\pi<\operatorname{Im}t<\pi\ .
\end{equation}
The boundaries of this strip do not, however, necessarily correspond to singularities of the analytically continued function. To establish the
continuation, we fix an angle $0<\theta<\pi/2$ and deform both large-frequency tails into the upper half-plane, leaving the finite part of the contour
unchanged. The deformed contour $\mathcal C_\theta$, oriented from its negative-frequency end to its positive-frequency end, approaches the rays
\begin{equation}
    \begin{aligned}
        \hat\omega&=-re^{-i\theta},
        &&\text{at the negative-frequency end},\\
        \hat\omega&=re^{i\theta},
        &&\text{at the positive-frequency end},
    \end{aligned}
    \qquad r\to\infty\ .
\end{equation}
This contour deformation is illustrated in Figure \ref{fig:exact-reflection-contour}.
The finite-frequency part remains unchanged, while both large-frequency tails are deformed into the upper half-plane without crossing the cut or any poles of the reflection coefficient \eqref{reflection coefficient}. Indeed, neither the $[-1,1]$ cut nor the poles on the imaginary axis are crossed.

To justify the deformation, we show that the closing arcs vanish at infinity. Applying Stirling's formula separately to each of the gamma functions in the exact reflection coefficient in these sectors gives
\begin{equation}\label{estimate}
        \log\left|R^{\rm exact}(\hat\omega)\right|
        =
        -\pi r\cos\phi-\frac{2r}{k}
        \left[
            \sin\phi
            \left(\log\frac{r}{k}-1\right)
            +\phi\cos\phi
        \right]
        +\mathcal O(\log r)\,,
\end{equation}
for either $\hat\omega=re^{i\phi}$ or
$\hat\omega=-re^{-i\phi}$, uniformly for $0\leq\phi\leq\theta$. In both sectors, the exact string correction therefore produces a negative contribution proportional to $r\sin\phi\log r$.

Having deformed the contour, we first establish agreement with the original integral inside its convergence strip,
\begin{equation}\label{agreement-old-new-contour}
    R(t) = \frac{1}{2\pi} \int_{\mathcal C_\theta} d\hat\omega\, e^{-i\hat\omega t} R^{\rm exact}(\hat\omega)\ , \qquad |\operatorname{Im}t|<\pi\ .
\end{equation}
To this end, truncate the old and new contours at $|\hat\omega|=r$ and join their corresponding endpoints by circular arcs. On either arc, for sufficiently large $r$, the Fourier kernel satisfies
\begin{equation}
    \left|
        e^{-i\hat\omega t}R^{\rm exact}(\hat\omega)
    \right|
    \leq
    \exp\left[
        -(\pi-|\operatorname{Im}t|)\,r\cos\theta
        +\mathcal O(\log r)
    \right]\ .
\end{equation}
For $|\operatorname{Im}t|<\pi$, this bound decreases exponentially with $r$. Since each arc has length proportional to $r$, both arc integrals vanish as $r\to\infty$. Cauchy's theorem therefore yields \eqref{agreement-old-new-contour}.

It remains to determine the domain of convergence of the new representation \eqref{agreement-old-new-contour}. Along either asymptotic ray, \eqref{estimate} implies that, for $t$ in any compact set $K\subset\mathbb C$,
\begin{equation}\label{entire-bound}
    \left|
        e^{-i\hat\omega t}R^{\rm exact}(\hat\omega)
    \right|
    \leq
    \exp\left[
        -\frac{2\sin\theta}{k}\,r\log r
        +C_{\theta,K}r
    \right]\,,
\end{equation}
for all sufficiently large $r$, where $C_{\theta,K}$ is independent of $r$. Because $\theta>0$, the negative $r\log r$ term dominates every term linear in $r$. Both tails therefore converge absolutely and uniformly on every compact subset of the complex $t$-plane. Differentiating with respect to $t$ introduces only powers of $\hat\omega$, which do not spoil this convergence. Together with the unchanged compact part of the contour, the rotated integral thus defines an entire function of $t$.

Since this representation agrees with the original Fourier integral throughout the strip $|\operatorname{Im}t|<\pi$, it provides its unique entire continuation. In particular, $t=i\pi$ is regular, and the pole found in the classical analysis is absent.

\begin{figure}[t]
    \centering
    \includegraphics[width=\linewidth]{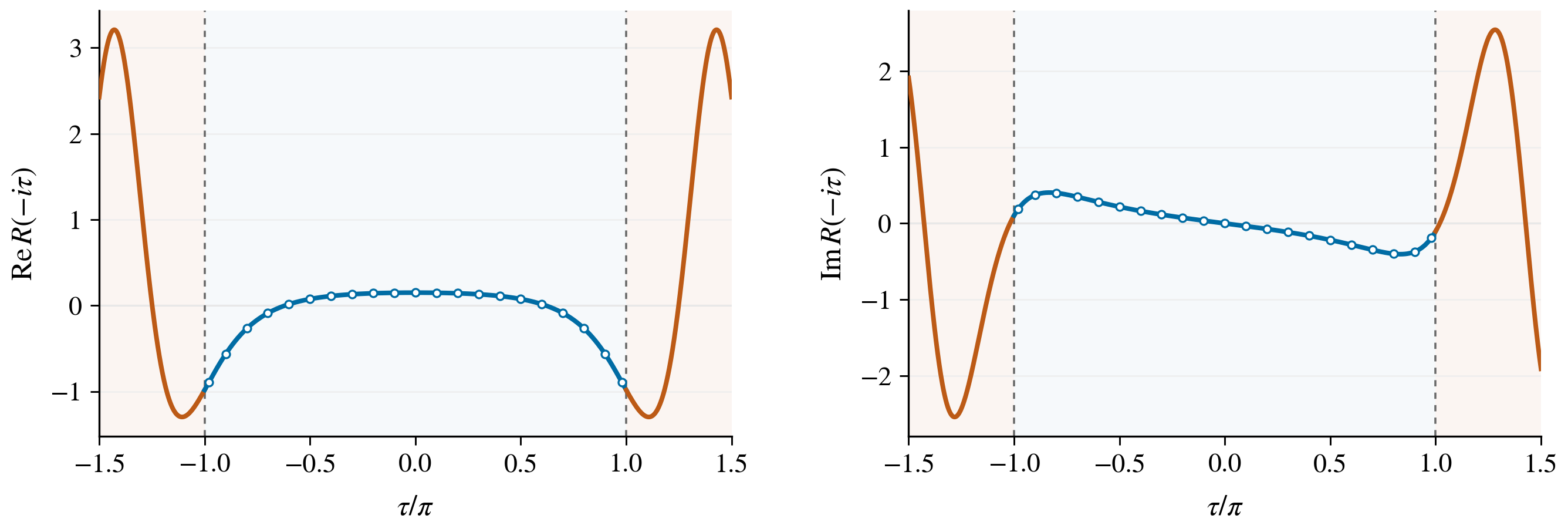}
    \caption{The real (left) and imaginary (right) parts of the exact two-sided Fourier transform $R(-i\tau)$, for $k=2$ and $\nu(k)=1$. The solid curves are evaluated using the deformed contour. Blue indicates $|\tau|<\pi$, where the original real-frequency integral converges absolutely, and orange indicates its analytic continuation beyond this domain. Circles show independent evaluations of the original real-frequency integral.}
\label{fig:reflection-imaginary-time}
\end{figure}

For illustration, Figure \ref{fig:reflection-imaginary-time} shows the exact Fourier transform along the imaginary-time axis $\tau=it$, for $k=2$ and $\nu(k)=1$. We evaluate the full gamma-function expression on the deformed contour at each sampled value of $\tau$. Inside $|\tau|<\pi$, these values agree with independent evaluations of the original real-frequency integral, providing a numerical check of the contour deformation. The transform remains finite and smooth across $\tau=\pm\pi$, and the
classical pole at $t=i\pi$ is absent.

\subsection{Large-\texorpdfstring{$k$}{k} expansion near the bouncing singularity}

Here we discuss the reflection amplitude in large-$k$ perturbation theory. Recall that the exact amplitude is related to the classical one by \eqref{quantum-classical-reflection-relation}. Correctness of the classical limit requires $\nu(k)\to1$ as $k\to\infty$. We shall assume that $\nu(k)$ admits an expansion of the form
\begin{equation}
    \log\nu(k)
    =
    \frac{c_1}{k}+\frac{c_2}{k^2}
    +\frac{c_3}{k^3}+\cdots\ .
\end{equation}
At fixed $\hat q$, expanding the logarithm of the ratio of gamma functions in \eqref{quantum-classical-reflection-relation} gives
\begin{equation}
    R^{\rm exact}(\hat\omega)=
    \left[ 1-\frac{i\hat q(c_1+2\gamma_E)}{k} + \frac{-i\hat q c_2-\frac12\hat q^2(c_1+2\gamma_E)^2}{k^2} + O(k^{-3})\right]
    R_{\rm classical}^{\rm exact}(\hat\omega)\,,
\end{equation}
where $\gamma_E$ is the Euler-Mascheroni constant. This expansion retains the dispersion relation $\hat q^2=\hat\omega^2-1$ following from
\eqref{simplified-physical-state-condition}. To extract the singular behavior of the Fourier transform, we use the positive-frequency asymptotics
\eqref{R-cigar-classical-leading},
\begin{equation}
    R_{\rm classical}^{\rm exact}(\hat\omega)
    \sim -2i e^{-\pi\hat\omega}\,,
    \qquad \hat q\sim\hat\omega\ .
\end{equation}
The Fourier kernel combines with the exponential falloff to give $e^{-i\hat\omega(t-i\pi)}$. Using
\begin{equation}
    \int_0^\infty d\hat\omega\,
    e^{-i\hat\omega(t-i\pi)}
    =
    \frac{1}{i(t-i\pi)}\,,
    \qquad \operatorname{Im}t<\pi\,,
\end{equation}
and obtaining the higher frequency moments by applying $i\partial_t$, we find, within this high-frequency approximation,
\begin{equation}
    R(t) \simeq -\frac{1}{\pi(t-i\pi)} +\frac{c_1+2\gamma_E}{\pi k(t-i\pi)^2} +\frac{1}{\pi k^2}
      \left(\frac{c_2}{(t-i\pi)^2} -\frac{(c_1+2\gamma_E)^2}{(t-i\pi)^3} \right) + O(k^{-3})\ .
\end{equation}
The higher poles generated by the large-$k$ expansion show that the singularity resolution is a non-perturbative effect in $k$.

\section{Discussion}
\label{S:Discussion}

In this paper, we analyzed the reflection amplitude and its position-space singularities for several asymptotically flat spaces. Working with the Schwarzschild black hole and the cigar geometry, which is a part of the NS5 supergravity background, we related the leading singularity in the complex time plane to a two-leg null ‘bouncing' geodesic. The cigar case arises as the classical $k\to\infty$ limit of the $SL(2,\mathbb{R})_k/U(1)$ conformal field theory, for which the reflection amplitude is known exactly. We showed that, for any finite $k$, the latter amplitude can be analytically continued to an entire function in the complex--$t$ plane. Therefore, the ‘bouncing singularity' gets resolved by worldsheet effects.

The large momentum behavior of the reflection coefficient  responsible for the resolution of the bouncing singularity found in this paper was also used in \cite{Ben-Israel:2017zyi,Itzhaki:2018rld} to study the effective potential behind the horizon. It was argued that string corrections produce singular behavior just inside the horizon. The resulting singularities of the potential should be distinguished from the bouncing singularities considered here. Our calculation concerns an observable defined in the asymptotic region and does not require reconstructing an interior potential.

The same large momentum behavior of the stringy reflection coefficient was studied in \cite{Giveon:2015cma,Ben-Israel:2015mda}. In \cite{Giveon:2015cma}, the $p\log p$ phase was obtained from the radial zero-mode integral in the sine-Liouville theory and interpreted in terms of the winding condensate. 
A Lorentzian interpretation of the winding condensate was proposed in \cite{Jafferis:2021ywg}. In this description, the eternal black hole in the Hartle--Hawking state is dual to two disconnected linear-dilaton backgrounds in an entangled state, with a condensate of entangled folded strings. See also \cite{Giveon:2016dxe} for the identification between geometric and winding modes. It would be interesting to understand the resolution of the bouncing singularity in this description.

Our result also suggests that string corrections remove the bouncing singularity of the retarded bulk Green's function. The relation between bouncing geodesics and singularities of bulk Green's functions was studied in \cite{Grozdanov:2026cut}, and extended to asymptotically flat and de Sitter black holes in \cite{Grozdanov:2026ktq}. In particular, this relation can be discussed directly in the bulk, without referring to a boundary field theory.

For two insertions in the asymptotic region of the NS5 background, the canonically normalized radial Green's function takes the form\footnote{This expression is valid away from the upper-half-plane poles of $R$, noticed in \cite{Parnachev:2005hh}. In the full classical Green's function, these poles cancel against terms omitted here. The cancellation is shown in Appendix~\ref{app:cigar-green-function}.}
\begin{equation}\label{discussion-Green-function}
    G_R(\hat\omega;z,z')
    \stackrel{z,z'\to\infty}{\simeq}
    \frac{1}{2i\hat q}
    \left[ e^{i\hat q|z-z'|}+R(\hat\omega)e^{i\hat q(z+z')} \right]\,,
    \qquad \hat q=\sqrt{\hat\omega^2-1}\,,
\end{equation}
at zero spatial momentum along the branes. This is the NS5 counterpart of the Schwarzschild relation in eq. (1.2) of \cite{Correia:2025enx}. The direct term is regular at the bouncing time. The reflected term contains the same scattering amplitude studied here, multiplied by the external propagation factor. 

The frequency-space analytic structures of $R$ and $G_R$ nevertheless differ. In Schwarzschild, Stokes phenomena give the reflection amplitude a cut along the positive imaginary-frequency axis which is absent from the finite-radius retarded Green's function \cite{Correia:2025enx}. This does not contradict the Schwarzschild bouncing singularity discussed in Section~\ref{S:Reflection amplitudes and bouncing geodesics}. Its origin is the large positive real-frequency behavior
\begin{equation}
    R_\ell(\omega) \simeq C_\ell e^{-\beta\omega/2}\,,
\end{equation}
which gives a singularity of the reflection amplitude at $t=i\beta/2$. 

At finite $k$, the additional stringy phase changes the large-frequency behavior responsible for this singularity. On the contours used in Section~3.3, the exact reflection amplitude decays faster than exponentially. The external propagation phases and inverse powers of frequency cannot compensate this decay, so the reflected contribution remains regular at the classical bouncing time. Hence, the resolution found here will persist in the analytically continued retarded bulk Green's function.

Bouncing singularities are often associated with asymptotic lines of quasinormal modes in frequency space. In the NS5 example, however, the classical pole locations remain unchanged at finite $k$, even though the bouncing singularity disappears.  In the classical theory, the leading large-$n$ terms give (upon closing the contour in the lower half $\hat\omega$-plane and picking up the contributions of the quasinormal modes)
\begin{equation}\label{discussion-classical-pole-sum}
    R_0(t)\sim t \sum_{n\geq2}(-1)^n e^{-(n+{1\over2}) t} , \qquad \operatorname{Re} t>0\,,
\end{equation}
whose analytic continuation has a pole at $t=i\pi$. At finite $k$, both the residues and the arc contributions get enhanced by $\exp(i \hat \omega \log \hat \omega)$, with the result being the complete resolution of the singularity.

There are several open questions to be addressed in the future. Firstly, it would be good to understand how generally this singularity resolution occurs. While our example shows that nonperturbative effects at finite string length can resolve the singularity, it remains to be seen whether similar physics occurs in, for example, $\mathcal{N}=4$ super Yang--Mills theory as the 't Hooft coupling is lowered from infinity.

Another possible direction concerns cosmological applications. The $SL(2)/U(1)$ coset has also been used to construct string backgrounds with cosmological singularities \cite{Kounnas:1992wc,Craps:2002ii,Buchel:2002kj,Berkooz:2002je,Giveon:2003gb,Toumbas:2004fe,Kounnas:2007pg}. It would be interesting to investigate whether the mechanism discussed here has an analogue in these time-dependent backgrounds, and whether studying the corresponding string amplitudes could shed light on the fate of cosmological singularities.

Finally, it would be interesting to explore the implications of our results for higher-dimensional black holes in the large-dimension limit. The appearance of the $SL(2)/U(1)$ coset in this limit \cite{Emparan:2013xia,Chen:2021emg} suggests a possible connection between the stringy resolution found here and the fate of bouncing singularities in more general black-hole backgrounds.

\section*{Acknowledgements}

We wish to thank N. \v Ceplak,  M. Kulaxizi, H. Liu, G. Policastro, M. Rangamani, S. Valach for discussions. This publication has emanated from research conducted with the financial support of Taighde Éireann – Research Ireland under Grant number SFI-22/FFP-P/11444. AP thanks the Aspen Center for Physics, which is supported by National Science Foundation grant PHY-2210452, and the Henri Poincaré Institute, Paris (IHP), for hospitality during the work on this project.

\appendix

\section{Bulk Green function in the cigar geometry}\label{app:cigar-green-function}

In this appendix, we relate the classical reflection amplitude to the retarded bulk Green function at zero spatial momentum along the fivebranes. It is convenient to introduce the following notation
\begin{equation} \label{eq:app-parameters}
 a=\frac{1+i\hat q-i\hat\omega}{2}\,, \qquad b=\frac{1-i\hat q-i\hat\omega}{2}\,,
\end{equation}
where the branch of $\hat q=\sqrt{\hat\omega^2-1}$ is fixed by $\hat q\sim\hat\omega$ at infinity. The solutions which are incoming and outgoing at infinity are
\begin{align}
 f_-(\hat\omega,z) &=y^{i\hat q/2}(1-y)^{-i\hat\omega/2} {}_2F_1(a,a;1+i\hat q;y)\,,\nonumber\\
 f_+(\hat\omega,z)&=y^{-i\hat q/2}(1-y)^{-i\hat\omega/2} {}_2F_1(b,b;1-i\hat q;y) \label{eq:app-modes}
\end{align}
with $y=(1+e^{2z})^{-1}$. They are normalized so that $f_\pm\simeq e^{\pm i\hat qz}$. The horizon-ingoing solution, normalized to unit incoming amplitude at infinity, is $f_-+Rf_+$, where throughout this appendix $R(\hat\omega) \equiv  R_{\rm classical}^{\rm exact}(\hat\omega)$. Since the Wronskian of the solutions reads $W(f_-,f_+)=2i\hat q$, the retarded Green function is
\begin{equation}\label{eq:app-green}
 G_R(\hat\omega;z,z') = \frac{[f_-(z_<)+R(\hat\omega)f_+(z_<)]f_+(z_>)}{2i\hat q}\,,
\end{equation}
where $z_<=\min(z,z')$ and $z_>=\max(z,z')$.
The two factors impose the horizon and infinity boundary conditions; the Wronskian in the denominator gives a unit jump in $\partial_zG_R$ at $z=z'$. \eqref{eq:app-green} is exact at finite radius. Replacing the radial solutions by plane waves gives the asymptotic formula \eqref{discussion-Green-function} used in the discussion section.

\subsection*{Cancellation of the upper-half-plane poles}

The reflection coefficient has poles in the $\hat\omega$--plane at
\begin{equation}\label{eq:app-upper-poles}
 \Omega_m=i\lambda_m\,, \qquad \lambda_m = \sqrt{m^2-1}\,, \qquad \hat q(\Omega_m)=im\,, \qquad m=2,3,\ldots\ .
\end{equation}
We will compute the singular terms in both $R$ and $f_-$ at one of these poles. Introduce the local variable in the vicinity of $\Omega_m$
\begin{equation}\label{eq:app-eta}
    \eta = i\hat q + m =\frac{i\lambda_m}{m}(\hat\omega-\Omega_m) + O((\hat\omega-\Omega_m)^2)\ .
\end{equation}
With this definition, a $1/\eta$ term corresponds to a simple pole in frequency. At the pole $\Omega_m$, we have
\begin{equation}\label{eq:app-pole-parameters}
 a_m=\frac{1-m+\lambda_m}{2}\,, \qquad b_m=a_m+m \ .
\end{equation}
At this point, the only singular term in $R_{\rm classical}^{\rm exact}(\hat\omega)$ is $\Gamma(i\hat q)=\Gamma(-m+\eta)$. Using $\Gamma(-m+\eta)=(-1)^m/(m!\eta)+O(1)$, we find
\begin{equation}\label{eq:app-R-pole}
 R(\hat\omega)=\frac{(-1)^m K_m}{\eta}+O(1)\,, \qquad K_m=\frac{\Gamma(b_m)^2}{m!(m-1)!\,\Gamma(a_m)^2} = \frac{(a_m)_m^2}{m!(m-1)!}\,,
\end{equation}
where $(a)_j=\Gamma(a+j)/\Gamma(a)$ denotes the Pochhammer symbol.
\smallskip

Now consider the hypergeometric function appearing in $f_-$. Its third parameter is $1+i\hat q=1-m+\eta$. Using properties of hypergeometric functions, we have
\begin{equation}\label{eq:app-hypergeom-pole}
 {}_2F_1(a,a;1-m+\eta;y) =\frac{(-1)^{m-1}K_m}{\eta}\, y^m{}_2F_1(b_m,b_m;m+1;y)+O(1)\,,
\end{equation}
as $\eta\to0$. Accounting for the prefactor in $f_-$, we obtain
\begin{equation}\label{eq:app-incoming-pole}
 f_-(\hat\omega,z) =\frac{(-1)^{m-1}K_m}{\eta}\,f_+(\Omega_m,z)+O(1)\ .
\end{equation}
Comparing with \eqref{eq:app-R-pole}, the coefficients have opposite signs and the same magnitude. Consequently,
\begin{equation}\label{eq:app-cancellation}
 f_-(\hat\omega,z)+R(\hat\omega)f_+(\hat\omega,z) =\frac{[(-1)^{m-1}+(-1)^m]K_m}{\eta}\, f_+(\Omega_m,z)+O(1) = O(1)\ .
\end{equation}
Equivalently, the frequency residues are
\begin{equation}
 r_m\equiv\underset{\hat\omega=\Omega_m}{\operatorname{Res}}R
 =\frac{(-1)^m mK_m}{i\lambda_m}\,,\qquad
 \underset{\hat\omega=\Omega_m}{\operatorname{Res}}f_-(\hat\omega,z)
 =-r_m f_+(\Omega_m,z)\ .
 \label{eq:app-residues}
\end{equation}
The remaining factor $f_+(z_>)/(2i\hat q)$ in \eqref{eq:app-green} is regular: $\hat q=im\ne0$, and the third parameter of its hypergeometric function is $1+m$. This proves that $G_R$ has no pole at $\Omega_m$ for any finite real $z,z'$. Replacing the radial solutions by plane waves discards the pole in $f_-$ and hence misses the cancellation.

\subsection*{The bouncing singularity}

The bouncing singularity can instead be recovered from the lower-half-plane poles of the full Green's function. They occur at $b=-n$, giving
\begin{equation}\label{eq:app-qnm}
 \hat\omega_n = -\frac{i}{2} \left(2n+1-\frac{1}{2n+1}\right)\,, \qquad n=1,2,\ldots\ .
\end{equation}
These poles are of order two. We write
\begin{equation}\label{eq:app-laurent}
 G_R(\hat\omega;z,z') =\frac{D_{2,n}}{(\hat\omega-\hat\omega_n)^2} +\frac{D_{1,n}}{\hat\omega-\hat\omega_n}+O(1)\ .
\end{equation}
For fixed $z,z'>0$, let $Z=z+z'$. Expanding the gamma functions in $R$, and using $f_+(\hat\omega_n,z)\sim e^{(n+1/2)z}$, gives
\begin{equation}\label{eq:app-coefficients}
 D_{2,n}=\frac{(-1)^n e^{(n+1/2)Z}}{2\pi^2n} \left[1+O\!\left(\frac{\log n}{n}\right)\right]\,, \qquad \frac{D_{1,n}}{D_{2,n}}=iZ+O(n^{-1})\ .
\end{equation}
Closing the frequency contour in the lower half-plane, initially at real $t>Z$, each double pole contributes $-i e^{-i\hat\omega_nt}(D_{1,n}-itD_{2,n})$. With $\Delta t=t-Z$, the leading large-$n$ sum is therefore
\begin{equation}\label{eq:app-sum}
 \begin{split}
 G_R(t;z,z')\big|_{\rm leading}
 &=-\frac{\Delta t\,e^{-\Delta t/2}}{2\pi^2}
 \sum_{n=1}^{\infty}\frac{(-1)^n e^{-n\Delta t}}{n}\\
 &=\frac{\Delta t\,e^{-\Delta t/2}}{2\pi^2}
 \log(1+e^{-\Delta t})\ .
 \end{split}
\end{equation}
The subleading terms are less singular. The threshold cut has bounded frequency support and does not contribute a singularity at a finite complex time.

Continuing \eqref{eq:app-sum} from $\operatorname{Re}(\Delta t)>0$, we find
\begin{equation}
 G_R(t;z,z')=\frac{1}{2\pi}\log(t-t_c)
 +\text{less singular terms}\,,
 \qquad t_c=z+z'+i\pi\ .
 \label{eq:app-bounce}
\end{equation}
There is a conjugate singularity at $z+z'-i\pi$. To compare these times with the null trajectory, recall that the radial metric is conformal to $-dt^2+dz^2$. An ingoing null segment obeys $dt=-dz$, whereas an outgoing segment obeys $dt=dz$. Place the source at $(t,z)=(0,z')$. The ingoing segment ends at the singularity at tortoise coordinate $z_s$, after a time $z'-z_s$. The outgoing segment from $z_s$ to the observation point $z$ takes a further time $z-z_s$. Adding the two gives
\begin{equation}\label{eq:app-null-time}
 t_c=\int_{z'}^{z_s}(-dz'') + \int_{z_s}^z dz'' =(z'-z_s)+(z-z_s)=z+z'-2z_s\ .
\end{equation}
The bulk Green's function thus detects the same classical bounce as the reflection amplitude.

\section{Analytic structure of the classical reflection coefficient}
\label{app:sugra-analytic-structure}

\begin{figure}[htbp]
    \centering
    \includegraphics[width=0.85\linewidth]{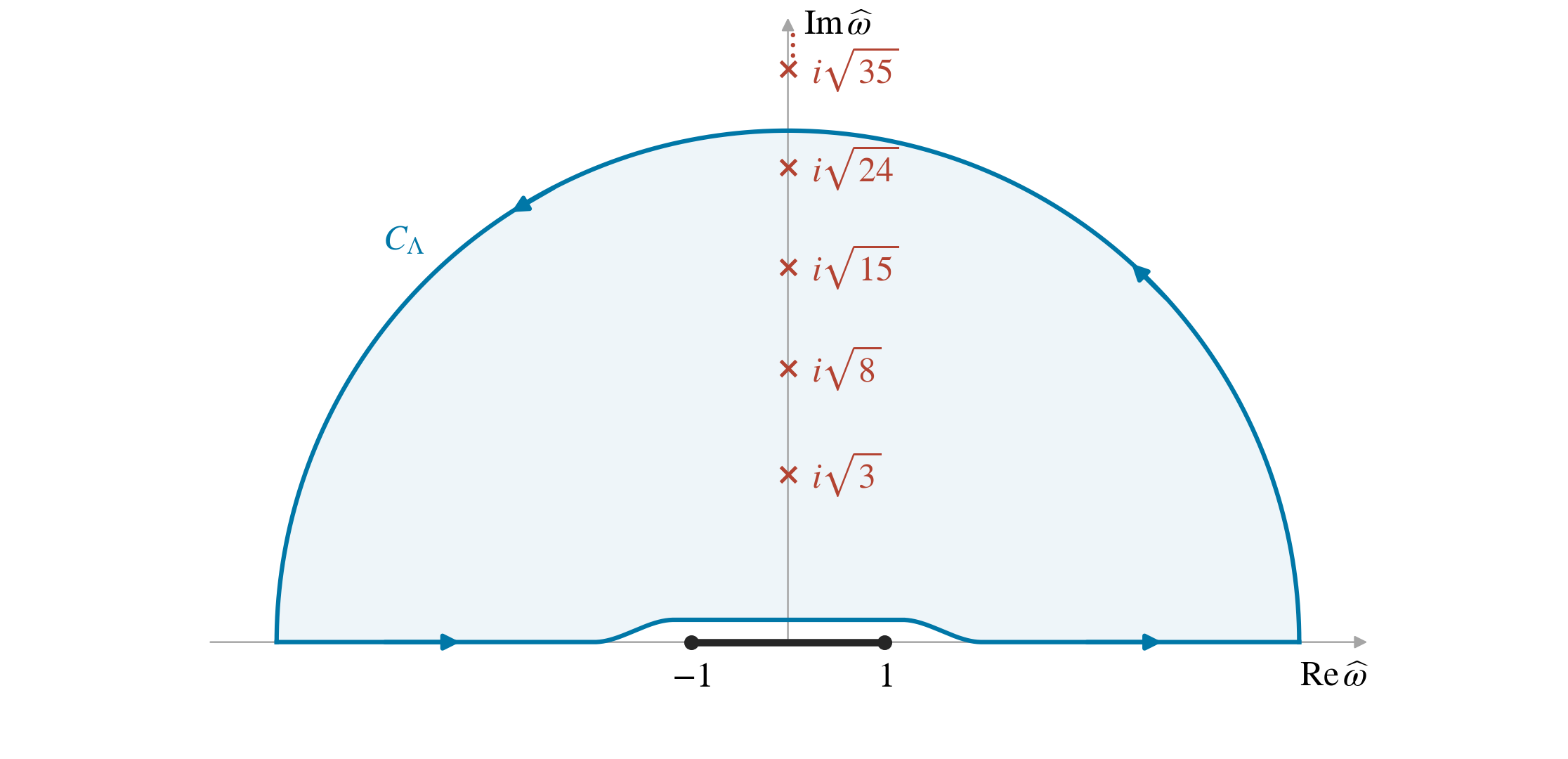}
    \caption{Frequency contour used to evaluate the supergravity reflection amplitude $R(t)$ for $t<0$. The contour follows the real axis, passes above the cut $[-1,1]$, and closes counterclockwise along the upper arc $C_\Lambda$. Red crosses mark the simple poles. The cut remains outside the enclosed region.}
    \label{fig:sugra-frequency-contour}
\end{figure}

\begin{figure}[htbp]
    \centering
    \includegraphics[width=0.85\linewidth]{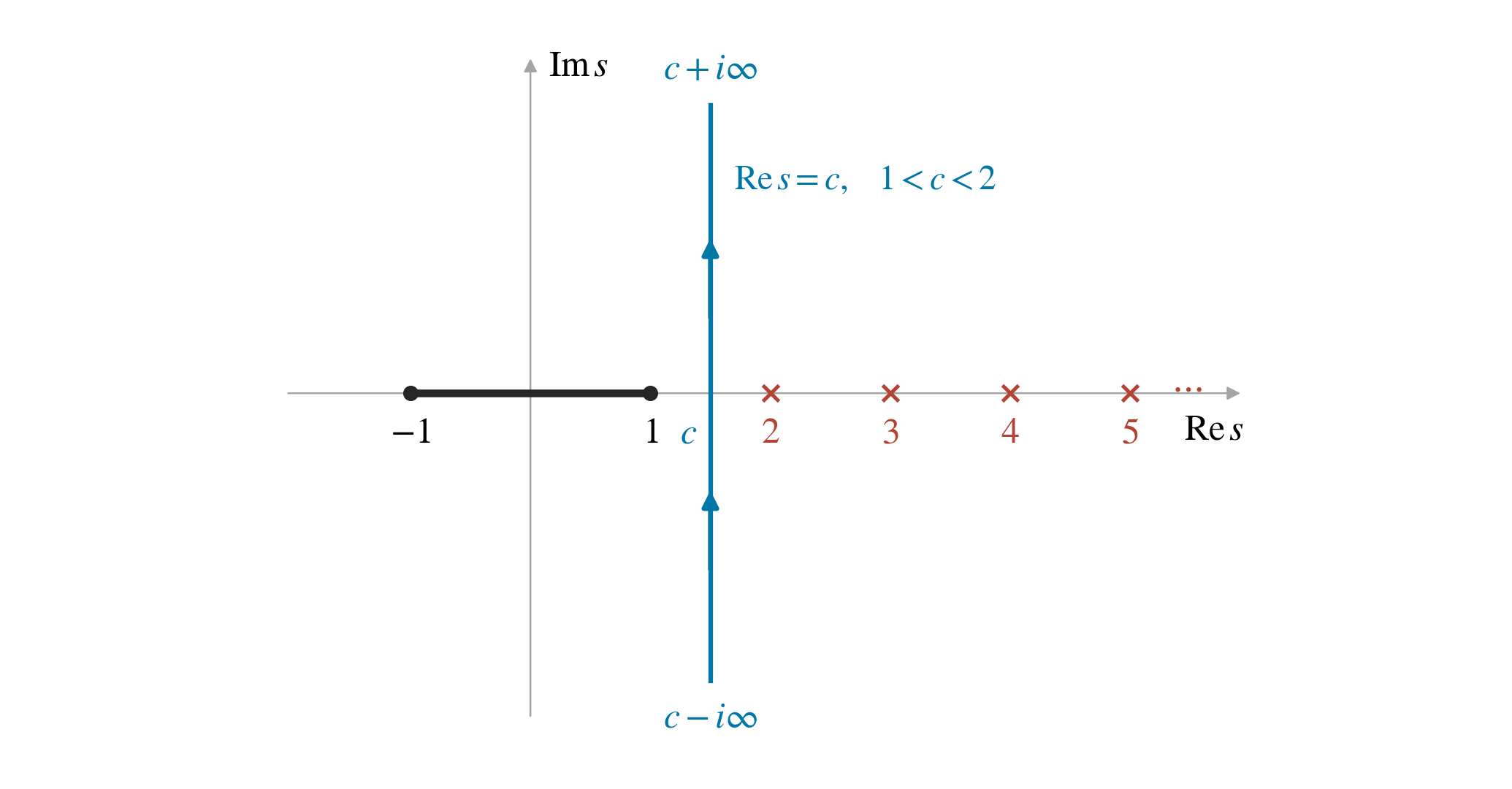}
    \caption{Auxiliary contour for the analytic continuation of $R(t)$. The same contour is used for every $m$, which labels the convergence strip in the complex-time plane.}
    \label{fig:sugra-auxiliary-contour}
\end{figure}

In this appendix, we determine the complex-time singularities of the exact supergravity amplitude \eqref{R-cigar-classical-exact}, extending the leading-frequency argument that gives \eqref{bouncing-singularity-cigar}.  Note that this calculation does not apply to the retarded Green function, which has no poles in the upper half-plane (see Appendix \ref{app:cigar-green-function} for the details). We use the same frequency sheet and Fourier prescription as for the string-corrected
amplitude,
\begin{equation}
 \hat q=\sqrt{\hat\omega^2-1}\sim\hat\omega,
 \qquad
 R(t)=\frac1{2\pi}\int_{-\infty}^\infty d\hat\omega\, e^{-i\hat\omega t}R_{\rm classical}^{\rm exact}(\hat\omega+i\epsilon)\ .
 \label{app:sugra-transform}
\end{equation}
The amplitude $R(\hat\omega)$ has a cut along the interval $[-1,1]$. We compute the Fourier transform first for real $t<0$, where the contour closes in the upper half-plane, as shown in Figure \ref{fig:sugra-frequency-contour}, and then analytically continue the result.
For $t<0$, the factor $e^{-i\hat\omega t}$ decays in the upper half-plane. The reflection formula for the $\Gamma$-function gives a useful equivalent expression for \eqref{R-cigar-classical-exact},
\begin{equation}
R_{\rm classical}^{\rm exact}(\hat\omega) = \frac{\pi}{\hat q\,\sinh(\pi\hat q)}
\left[\frac{\Gamma\!\left(\frac{1-i\hat q-i\hat\omega}{2}\right)}{\Gamma(-i\hat q)\,\Gamma\!\left(\frac{1+i\hat q-i\hat\omega}{2}\right)}\right]^2\ .
\end{equation}
Stirling estimates allow the large arcs to be discarded along contours avoiding the poles. Since the original contour passes above the cut, the cut lies outside the enclosed region. The enclosed poles come from $\Gamma(i\hat q)$,
\begin{equation}
i\hat q=-n\ \Rightarrow\ \hat\omega=i\sqrt{n^2-1}\,,\qquad n\ge2\ .
\end{equation}
The remaining apparent pole for $n=1$ with $\hat\omega=0$ cancels against a zero of the remaining factors to give
\begin{equation}
R_{\rm classical}^{\rm exact}(0+i\epsilon)=\frac12\ .
\end{equation}
The counterclockwise contour contributes $2\pi i$ times the residues
\begin{equation}
\underset{\hat\omega=i\sqrt{n^2-1}}{\operatorname{Res}}\,
R_{\rm classical}^{\rm exact}(\hat\omega)
=
\frac{(-1)^n}{i\sqrt{n^2-1}}
\left[\frac{\Gamma\!\left(\frac{1+n+\sqrt{n^2-1}}{2}\right)}
{\Gamma(n)\, \Gamma\!\left(\frac{1-n+\sqrt{n^2-1}}{2}\right)}\right]^2\ .
\end{equation}
Including the Fourier normalization, we get the exact result
\begin{equation}\label{app:sugra-exact-series}
R(t)=
\sum_{n=2}^\infty \frac{(-1)^n}{\sqrt{n^2-1}}
\left[\frac{\Gamma\!\left(\frac{1+n+\sqrt{n^2-1}}{2}\right)}{\Gamma(n)\, \Gamma\!\left(\frac{1-n+\sqrt{n^2-1}}{2}\right)}\right]^2 e^{\sqrt{n^2-1}\,t}\,,
\qquad \operatorname{Re}t<0\ .
\end{equation}
Since $\sqrt{n^2-1}=n+O(n^{-1})$ and the coefficient approaches $(-1)^n/\pi$, this series converges absolutely and locally uniformly in the left half-plane. There, it agrees with the Fourier integral in their common domain of convergence
\begin{equation}
\operatorname{Re}t<0\,, \qquad |\operatorname{Im}t|<\pi\ .
\end{equation}
The original Fourier integral is holomorphic throughout $|\operatorname{Im}t|<\pi$, so it supplies the continuation of \eqref{app:sugra-exact-series} through $t=0$. The residue series itself should not be used directly for $\operatorname{Re}t\ge0$. To continue $R(t)$ beyond the left half-plane, we replace the sum by a contour integral,
\begin{equation}
\begin{aligned}
R(t) \equiv -\frac1{2i} \int_{c-i\infty}^{c+i\infty}ds\,
&\frac{e^{\sqrt{s^2-1}\,t-2\pi ims}}
{\sqrt{s^2-1}\,\sin\pi s}\\
&\times
\left[
\frac{\Gamma\!\left(\frac{1+s+\sqrt{s^2-1}}2\right)}
{\Gamma(s)\,
 \Gamma\!\left(\frac{1-s+\sqrt{s^2-1}}2\right)}
\right]^2\,,
\qquad
|\operatorname{Im}t-2\pi m|<\pi\ .
\end{aligned}
\label{app:sugra-continuation}
\end{equation}
Here $\sqrt{s^2-1}\sim s$, with the cut $[-1,1]$, and $1<c<2$. Thus the vertical contour lies to the right of the cut and to the left of the first integer pole contributing to the series, as shown in
Figure \ref{fig:sugra-auxiliary-contour}. The integer $m$ labels the convergence strip of this representation.
For $\operatorname{Re}t<0$, closing the contour to the right
reproduces \eqref{app:sugra-exact-series}. Indeed,
\begin{equation}
\underset{s=n}{\operatorname{Res}}\frac1{\sin\pi s} =\frac{(-1)^n}{\pi}\,, \qquad e^{-2\pi imn}=1\ .
\end{equation}
The clockwise orientation accounts for the prefactor $-1/(2i)$. On the vertical contour, Stirling estimates show that the
$\Gamma$-function factor, including $1/\sqrt{s^2-1}$, is bounded, while
\begin{equation}
\sqrt{s^2-1}=s+O(s^{-1})\,,
\qquad
\frac1{\sin\pi s}
=O\!\left(e^{-\pi|\operatorname{Im}s|}\right)\ .
\end{equation}
For each integer $m$, the contour integral \eqref{app:sugra-continuation} continues $R(t)$ into the horizontal strip $|\operatorname{Im}t-2\pi m|<\pi$. Together, these contour formulas and the residue series in the left half-plane give a single-valued analytic continuation of $R(t)$ to
\begin{equation}
\mathbb C\setminus \bigcup_{m\in\mathbb Z}\left\{t\in\mathbb C:\operatorname{Re}t\ge0,\quad\operatorname{Im}t=\pi(2m+1)\right\}\ .
\end{equation}
The excluded horizontal rays are chosen as branch cuts, with endpoints at $t=i\pi(2m+1)$. To determine the singularities at the endpoints, we consider the large-$n$ expansion of the summand in \eqref{app:sugra-exact-series},
\begin{equation}
\begin{aligned}
&\frac{(-1)^n}{\sqrt{n^2-1}}
\left[\frac{\Gamma\!\left(\frac{1+n+\sqrt{n^2-1}}{2}\right)}{\Gamma(n)\, \Gamma\!\left(\frac{1-n+\sqrt{n^2-1}}{2}\right)}\right]^2 e^{\sqrt{n^2-1}\,t}
\\[0.5em]
&\qquad =
\frac{(-e^t)^n}{\pi}\left[1-\frac{\log n+\frac12+\gamma_E+2\log2+t}{2n}+O\!\left(\frac{\log^2n}{n^2}\right)\right]\ .
\end{aligned}
\end{equation}
The leading terms sum to
\begin{equation}
\frac1\pi\sum_{n=2}^\infty(-e^t)^n =\frac{e^{2t}}{\pi(1+e^t)}\,,
\end{equation}
which has simple poles at $t=i\pi(2m+1)$. For $m=0$, the leading pole reproduces the bouncing singularity in \eqref{bouncing-singularity-cigar}. The potential $V(z)=1-(1+e^{2z})^{-2}$ has poles at $z=i\pi(m+\tfrac12)$; twice their tortoise displacement gives the same time lattice. The locations repeat with imaginary period $i\beta$, although $R(t)$ itself is not periodic. These are the supergravity singularities to be compared with the entire continuation of the string-corrected reflection kernel.

\bibliographystyle{JHEP}
\bibliography{bibliography} 

\end{document}